\documentclass[prl,twocolumn,showpacs,floatfix]{revtex4}
\usepackage{graphicx,epsf}

\begin{document}

\def\tende#1{\,\vtop{\ialign{##\crcr\rightarrowfill\crcr
\noalign{\kern-1pt\nointerlineskip}
\hskip3.pt${\scriptstyle #1}$\hskip3.pt\crcr}}\,}

\title{Dynamics of topological defects in a spiral: a scenario for the spin-glass phase of cuprates}
\author{V.\ Juricic$^1$, L.\ Benfatto$^1$, A.\ O.\ Caldeira$^2$, and C.\ Morais Smith$^1$}

\affiliation{$^1$D\'epartement de Physique, Universit\'e de Fribourg, P\'erolles,  
CH-1700 Fribourg, Switzerland.\\
$^2$ Instituto de F\'\i sica Gleb Wataghin, Universidade Estadual de Campinas,
13083-970, Campinas, SP, Brazil}

\begin{abstract}
We propose that the dissipative dynamics of topological defects in a spiral
state is responsible for the transport properties in the spin-glass phase of
cuprates. Using the collective-coordinate method, we show that topological
defects are coupled to a bath of magnetic excitations.
By integrating out the bath degrees of freedom, we find that the dynamical
properties of the topological defects are dissipative. The calculated damping
matrix is related to the in-plane resistivity, which exhibits an anisotropy
and linear temperature dependence in agreement with experimental data.   
\end{abstract}
\maketitle

The discovery of four static incommensurate (IC) charge and spin peaks by 
neutron scattering experiments in Nd-doped
La$_{2-x}$Sr$_x$CuO$_4$ (LSCO) \cite{tran} confirmed the proposal that in
these materials the holes form vertical/horizontal charge stripes 
\cite{zaanen}, breaking the paradigm of homogeneous charge distribution. 
Moreover, the observation of IC
dynamical spin correlations in superconducting LSCO $(x > 0.05)$ at the same
wave vectors as observed in the Nd-doped compounds \cite{yama} placed the
stripe picture on a more solid footing and raised further questions concerning 
the coexistence of magnetism and superconductivity. Later, two IC elastic
magnetic peaks have been observed within the spin-glass (SG) phase of LSCO
$(0.02 < x < 0.05)$ \cite{NSSG}, and their interpretation in terms of diagonal
stripe formation appeared to be rather natural, given that the value
of the magnetic IC follows the same linear dependence on doping as observed 
within the superconducting regime \cite{NSSG}. 
However, no charge order has ever been measured in the
SG regime, and the picture of well-ordered stripes, very far apart from 
each other due to the very low doping concentration, seems quite improbable 
because disorder and frustration effects, which act to destabilize the 
stripes, are expected to dominate in this phase.

A different explanation of the two IC peaks in the SG phase is that these
may arise from the formation of a spiral magnetic order 
which breaks the translational symmetry in the spin 
but not in the charge sector \cite{nils}. In this model, the randomly 
distributed
holes act as frustration centers for the underlying anti-ferromagnetic (AF)
background, generating 
a dipole moment. A fraction of these dipoles may order ferromagnetically, while
the  others may remain disordered. If the number of ordered
dipoles  increases linearly with the doping concentration, which would appear
plausible, this model can describe the linear variation 
of the IC magnetic wave vector with doping 
\cite{NSSG}, without invoking  charge ordering \cite{nils}.

Recently, transport properties along both diagonal-directions in the Cu 
lattice have been investigated in high-quality, detwinned LSCO 
samples, and a rather strong anisotropy was detected 
\cite{ando}. However, this experiment provides no definitive evidence in favor
of the stripe picture, because, as we will show, a similar anisotropic 
behavior is expected for transport parallel or perpendicular to the spiral 
axis. 

In this Letter we investigate the formation of topological defects in the spiral
state that couple to excitations of the magnetic environment and diffuse
across the system. We assume that the charge carriers are attached 
to these defects and therefore the description of defect dynamics 
corresponds ultimately to the electrical transport properties
of cuprates in the SG phase. We evaluate the damping matrix, which is related
to the mobility of the topological defect, and show that its behavior at 
high temperatures follows the experimentally observed temperature ($T$) dependence of the
in-plane resistivity. In addition, we account for the anisotropy in the 
spin-wave
velocity in the directions parallel and perpendicular to the spiral axis, and
find that, in agreement with experiments, it leads to an anisotropy in the 
resistivity. 

The magnetic properties of charge transfer insulators, such as undoped
La$_2$CuO$_4$, are usually studied within the quantum non-linear sigma model
(NL$\sigma$M), which can correctly represent the long wavelength modes of
the Heisenberg model. This continuum model describes slow fluctuations of
the locally well defined staggered magnetization ${\bf n}$ (with 
${\bf n}^2 = 1$). When holes are added to the system, the magnetism becomes
more complex because those act simultaneously as a source of disorder and
dipolarlike frustration. The dipolar frustration can be described via
a coupling of the dipoles to the gradient of the order parameter ${\bf n}$
of the NL$\sigma$M. By accounting for the random distribution of dipolar
centers, the resulting Hamiltonian favors the formation of a
spiral phase, with a non-zero average twist $\partial_\mu {\bf n}$ of the
AF order parameter and a concomitant alignment of some of the 
dipoles \cite{nils}. 

The spiral ground-state breaks the O(3) spin symmetry completely. Its order
parameter, which belongs to the SO(3) group, is given in terms of an 
orthonormal basis ${\bf n}_k$, with $k = 1,2,3$ and $n_k^a n_q^a = 
\delta_{kq}$ \cite{apel}. The latter can be related to an element $g$
of SU(2) through $n_k^a =  (1/2) \mbox{tr} \left( \sigma^a
g  \sigma^k g^{-1} \right)$, where $\sigma^a$ are Pauli matrices. 
It is also convenient to introduce the fields 
$A_\mu^a = (1 /2i) \mbox{tr} \left( \sigma^a  g^{-1}
\partial_\mu g \right)$, which are related to the first derivatives 
of ${\bf n}_k$ through $\partial_\mu n_k^a = 2 \epsilon_{ijk} A_\mu^i n_j^a$
\cite{polyakov}.
Here $\mu$ stands for one-time and two-spatial coordinates. 
The Lagrangian describing the spiral state is given by 
\begin{equation}
\label{L1}
L  = {\cal N} \int  dx_\parallel dx_\perp
\left( {\bf A}_0^2-c_\perp^2 {\bf A}_\perp^2-
c_\parallel^2 {\bf A}_\parallel^2 \right),
\end{equation}
where ${\cal N} = 2 J S^2 /  c_\parallel c_\perp$, 
$J$ is the AF exchange and $c_\perp$, $c_\parallel$ are the 
spin wave velocities perpendicular and parallel to the spiral axis, with
$(c_\parallel/c_\perp)^2=\cos{(Qa)}$ \cite{nils}. Here $a$ is the lattice
constant and $Q$ is the IC magnetic wave vector observed by neutron 
scattering. Lagrangian (\ref{L1}) may be mapped to an isotropic form by 
introducing the coordinates ${\bf r}=(x, y)$ with 
$x=\sqrt{c_\parallel/c_\perp}x_\perp$,
$y=\sqrt{c_\perp/c_\parallel}x_\parallel$. We then find 
$L = {\cal N} \int d^2r \left( {\bf A}_0^2-c^2 {\bf A}_\alpha^2\right)$, where 
$\alpha = x,y$ and $c=\sqrt{c_\perp c_\parallel}$ is the isotropic spin-wave 
velocity. This rescaling leads to an anisotropic elementary cell, which 
will be neglected in what follows. Further comments on the anisotropy
effects are included in the conclusions.

An appropriate treatment of the spiral state has to take into account the
existence of topological defects. In principle, in $(2+1)$ space-time 
dimensions two types of static topological defects can exist: vortices and skyrmions. Skyrmionlike excitations are classified according
to the second homotopy group of the order parameter space, which in our
case is trivial, $\pi_2(SO(3))=0$. Thus, there are no skyrmions in the
spiral state and the only possible defects are $Z_2$-vortices, given
by the first homotopy group $\pi_1(SO(3))=Z_2$ \cite{dombre}. 
These defects originate from the chiral degeneracy of the spiral, which
can rotate clock- or counter-clock-wise \cite{kawamura}.  
Their description is provided by the elements $g = g_s(\Psi)
g_\varepsilon ({\vec \varepsilon})$ of the SU(2) group, with
$g_s = \exp{(i m^a  \sigma^a \Psi/2)}$ 
and $g_\varepsilon = \exp{(i \sigma^a  \varepsilon^a/2)}$.
The parameters ${\bf m}\Psi$ are related to the local spin, where ${\bf m}$ 
is a constant unit vector and $\Psi$ is a scalar field \cite{wintel}.
Our aim is to determine the topological defect solutions $\Psi_{2v}$ of
the saddle point equations of the spiral, as well as the vectorial field
${\vec \varepsilon}$ describing fluctuations around this solution. In terms
of these fields, the Lagrangian becomes 
\begin{eqnarray} 
L &=& {\cal N} \int  d^2 r \left(L_0 + L_1 \right), \qquad 
L_0 = \frac{1}{4} (\partial_\mu \Psi)^2 \\   \nonumber 
L_1 &=& \frac{1}{4} \left( \partial_\mu {\vec \varepsilon}\right)^2 + 
\frac{1}{2} \left \{ {\bf m} \cdot \left[ \partial_\mu {\vec \varepsilon} +
\frac{1}{2} \left( \partial_\mu {\vec \varepsilon} \times {\vec \varepsilon}
\right) \right] \right\}  \partial_\mu \Psi,
\end{eqnarray}
with $\partial_\mu A \partial_\mu B = \partial_t A \partial_t B - c^2 \nabla A
\nabla B$. This Lagrangian can be treated perturbatively because 
$|{\vec \varepsilon}|$
is small. 
We begin by considering the static limit of $L_0$, where the corresponding
equation of motion becomes the Laplace equation $\nabla^2\Psi({\bf{r}})=0.$ 
A nontrivial solution
$\Psi_{1v}({\bf r}, {\bf R})=\arctan(y-Y)/(x-X)$ has the form of a vortex 
centered at ${\bf{R}}=(X,Y)$, and its energy $E \propto \ln L/a$ diverges 
with the system size $L$. Unbound, free topological defects cannot, therefore, 
exist at low temperatures. However, the two-vortices solution 
$\Psi_{2v}= \Psi_{1v}({\bf r},{\bf R}_1)-\Psi_{1v}({\bf r},{\bf R}_2)$,
which describes a bound pair of topological defects (vortex and antivortex), 
centered respectively at ${\bf R}_1$ and ${\bf R}_2$, has a finite energy 
$E(\Psi_{2v})\propto \ln (d/a)$,
where $d =|{\bf{R}}_1-{\bf{R}}_2|$ is the modulus of the pair 
relative coordinate, which we take as a constant. Up to an irrelevant
constant, the two-vortices solution can be expressed in terms of the  
center of mass $ {\bf R} = ({\bf R}_1 + {\bf R}_2)/2$ and relative coordinate
${\bf d}$, yielding $\Psi_{2v}({\bf{r}})=\arctan \{ [({\bf r} - {\bf R}) 
\times {\bf d}]_{z}/[({\bf r} - {\bf R})^2- d^2/4]\}$. 

We will apply the collective coordinate method \cite{amir} to study the 
dynamics of a bound pair of topological defects in a spiral. 
The first task is to determine the dynamics of the fluctuations around the
defect. By evaluating $L_1 (\Psi_{2v})$ and expressing
the vectorial field in terms of polar coordinates, ${\vec \varepsilon} =
\left(\bar{\varepsilon} \cos \theta, 
\bar{\varepsilon} \sin \theta, \varepsilon^z \right)$, we find that
the $\varepsilon^z$ component is free and that $\theta = 
\Psi_{2v} /2$. The remaining equation of motion then reads 
$\left \{ \partial_t^2-c^2 [\nabla^2 + V({\bf r})] \right\} \bar{\varepsilon} =0$,
with $V({\bf r})= (\nabla\Psi_{2v})^2 / 4$. 
Its solution has the form $ \bar{\varepsilon} ({\bf{r}},t)= \sum_{nm}
q_{nm}(t)\eta_{nm}(\bf{r})$, 
where $\eta_{nm}$ are the eigenfunctions of the operator 
\begin{equation}
\label{etaeq}
- c^2 [\nabla^2 + V({\bf r})]\eta_{nm} = \omega^2_{nm} \eta_{nm}. 
\end{equation} 
The solutions of Eq.\ (\ref{etaeq}) have the form 
$\eta_{nm}= \sqrt{k_{nm}/8L} [\ H_{n}^{(1)}(k_{nm}r)+e^{-2 i
\delta_n}H_{n}^{(2)}(k_{nm}r)]\ e^{in\vartheta},$ 
where $H_{n}^{(1,2)}$  are Hankel functions of the first and 
second kinds, $\delta_n$ represents a phase shift for the mode $n$, 
and $\vartheta$ is a polar angle. 
The excitation with frequency
$\omega_{nm}$ has the dispersion  $\omega_{nm}= c |{\bf{k}}_{nm}|$. Equation\ (\ref{etaeq}) admits zero-frequency
modes associated with the continuous translational
symmetry of the model. A consistent treatment of them requires the use of collective coordinates. Thus, the center of the pair 
$\bf{R}$ is promoted to a dynamical variable ${\bf R} (t)$, namely   
$\Psi({\bf r}) \to  \Psi({\bf r} - {\bf R}(t))$ and
$\bar{\varepsilon}({\bf r},t) \to  \bar{\varepsilon}({\bf r} - {\bf R}(t), t) = 
\sum_{nm} q_{nm} (t) \eta_{nm}({\bf r}- {\bf R}(t))$. 

A lengthy but straightforward procedure \cite{vlad} leads to the effective 
Hamiltonian (in units of $2JS^2/c^2$) describing the pair of topological 
defects interacting with the bath of magnons,
\begin{equation}
\label{cham}
H=\frac{1}{2M}\left({\bf P}-{\bf P}_E \right)^2+\frac{1}{2}
\sum_{nm}\left(p_{nm}^2+\omega_{nm}^2\,q_{nm}^2\right), 
\end{equation}
where $M = \int d^2r (\nabla\Psi_{2v})^2 $ is the mass of the
topological defect, ${\bf P}$ and $p_{nm}$ are respectively the
momentum  conjugate to ${\bf R}$ and to the coordinate $q_{nm}$
of the excitation with frequency $\omega_{nm}$, and 
${\bf P}_E=\sum_{nm,kl}p_{nm}\,{\bf{G}}_{nm,kl}\,q_{kl}$.
The constants ${\bf G}_{nm,kl}$ are related to the eigenfunctions $\eta$ 
by ${\bf G}_{nm,kl}=\int d^2r \eta_{kl} \nabla\eta_{nm}$. 
The classical Hamiltonian (\ref{cham}) is quantized by introducing the usual
creation and annihilation operators $a^{\dagger}_{nm}$ and $a_{nm}$.
Restricting the analysis to the case of elastic scattering and retaining
only the terms at quadratic order in $a^{\dagger}_{nm}$, $a_{nm}$
we obtain
\begin{equation} 
\hat{H} = \frac{\hat{{\bf{P}}^2}}{2M}- \frac{\hbar \hat{\bf{P}}}{M} 
\sum_{nm,kl} {\bf D}_{nm,kl} a_{nm}a_{kl}^{\dagger}+\hat{H}_b, 
\end{equation}
where $\hat{H}_b=\sum_{nm}\hbar
\omega_{nm}a_{nm}^{\dagger}a_{nm}$ and  
${\bf D}_{nm,kl} = {\bf G}_{nm,kl}(\omega_{nm} + \omega_{kl})
/2i\sqrt{\omega_{nm}\omega_{kl}}$.

Our aim is to investigate, in the low-energy sector, the effective dynamics 
of the vortex-pair in the presence of the excitation bath.
We compute the reduced density matrix of the pair by integrating out the
degrees of freedom of the bath of magnons, for which we employ the 
Feynman-Vernon path-integral formalism. The evolution of the density matrix
for the full system is described by  $\hat{\rho}(t)= {\rm exp} (- i \hat{H}t /
\hbar) \hat{\rho}(0)  {\rm exp} (i \hat{H} t / \hbar)$. 
For the sake of simplicity, we
use the factorisable initial condition 
$\hat{\rho}(0)=\hat{\rho}_{v}(0)\,\hat{\rho}_{b}(0)$, 
where ${\rho}_{v}$ and $\hat{\rho}_{b}$ represent, respectively, the initial 
vortex-pair and bath density matrices. 
This condition implies that the pair of topological defects and the 
excitations do not interact at 
$t = 0$. The bath degrees of freedom are supposed to be initially in thermal
equilibrium, $ \hat{\rho}_{b}(0)= e^{-\beta\hat{H}_{b}} / 
{\rm tr} [e^{-\beta\hat{H}_{b}}]$, 
where $\beta \equiv \hbar/ k_{B}T$.
After evaluating the trace over these, we obtain the reduced
density-matrix operator for the vortex-pair 
$$ \hat{\tilde{\rho}}({\bf x},{\bf
y},t)=\int\ \int d{\bf x}'d{\bf y}'J({\bf x},{\bf y},t;{\bf x'},{\bf
y}',0)\hat{\rho}_{v}({\bf x}',{\bf y}',0), $$
where the super-propagator $J$ has the form
\begin{equation} 
\label{J}
J=\int_{{\bf x}'}^{\bf x}\mathcal{D}{{\bf x}}\int_{{\bf y}'}^{\bf y}
\mathcal{D}{\bf y}e^{\frac{i}{\hbar}[S_0[{\bf x}]-S_{0}[{\bf y}]]}
{\cal F}[{\bf{x}},{\bf{y}}].
\end{equation}
$S_{0}[{\bf{x}}]=\int_{0}^{t}dt'(M/2)\dot{{\bf{x}}}^2$ is the action
associated with the free motion of the vortices
and ${\cal F}[{\bf{x}},{\bf{y}}]$
is the influence functional, which describes the
effect of the excitations on the dynamics of the topological defect pair. 

Defining
the center of mass and relative coordinates  ${\bf{v}}=({\bf{x}}+{\bf{y}})/2,
{\bf{u}}=\bf{x}-\bf{y}$ and using the Born approximation, one finds,  
after a lengthy calculation \cite{vlad}, that  
\begin{equation}
\label{epsilonn}
{\cal F}[{\bf{x}},{\bf{y}}]=\exp{[i{\Phi}]}
\exp{[\tilde{\Phi}]},
\end{equation}
where 
\begin{widetext}
\begin{eqnarray} \nonumber
\Phi &=& 2\sum_{\mu,\nu=1}^2\int_{0}^{t}dt'\int_{0}^{t'}dt''
\epsilon^{\mu\nu}(t'-t'') \dot{u}^{\mu}(t')\dot{v}^{\nu}(t''), \qquad
\tilde{\Phi} = \sum_{\mu,\nu=1}^2 \int_{0}^{t}dt'\int_{0}^{t'}dt''
\tilde{\epsilon}^{\mu\nu}(t'-t'')\dot{u}^{\mu}(t') \dot{u}^{\nu}(t''), \\ \nonumber  
\epsilon^{\mu\nu}(t) &=& \sum_{nm, kl \neq 0} N_{nm}
D^{\mu}_{nm,kl} D^{\nu}_{nm,kl} \sin{(\omega_{nm}-\omega_{kl})t}, \qquad
{\tilde{\epsilon}}^{\mu\nu}(t) = \sum_{nm, kl \neq 0} N_{nm} D^{\mu}_{nm,kl}
D^{\nu}_{nm,kl} \cos{(\omega_{nm}-\omega_{kl})t} .
\end{eqnarray}
\end{widetext}
The boson occupation number is $N_{nm}=(e^{\beta\omega_{nm}}-1)^{-1}$. 
The effective action that describes the motion of the vortex pair in the 
presence 
of excitations may then be obtained by inserting Eq.\ (\ref {epsilonn})
into the superpropagator (\ref {J}). We find that
$S_{eff}=S_0[{\bf x}]-S_0[{\bf y}]+\hbar{\Phi}$ and the corresponding
equations of motion indicate that the dynamics of the defects is damped,
\begin{eqnarray} \nonumber &&
\ddot{v}^{\mu}(\tau)+\sum_{\nu}\int_{0}^{\tau}dt'
\gamma^{\mu\nu}(\tau-t')\,\dot{v}^{\nu}(t')=0, \\
&&
\ddot{u}^{\mu}(\tau)-\sum_{\nu}\int_{\tau}^{t}dt'\gamma^{\mu\nu}(t'-\tau)
\dot{u}^{\nu}(t')=0,
\label{em1}
\end{eqnarray}
where the matrix $\gamma^{\mu\nu}(t) = 
(2\hbar/M){\dot \epsilon}^{\mu\nu}(t)$ is a generalization of the 
damping coefficient. The decaying part $\exp[\tilde{\Phi}]$ of the
influence functional is related
to the diffusive properties of the defect, with  the diffusion matrix 
${\cal D}^{\mu\nu}(t)=\hbar {\ddot {\tilde{\epsilon}}}^{\mu\nu}(t)$.
The evaluation of the diffusion is analogous to that of the damping matrix, and
at low $T$ the two parameters are related by the fluctuation-dissipation
theorem. Here we will concentrate on $\gamma$ only.
We first introduce the scattering function 
$
S^{\mu\nu}(\omega,\omega')=  \sum
G^{\mu}_{nm,kl}G^{\nu}_{nm,kl}\delta(\omega-\omega_{nm})
\delta(\omega'-\omega_{kl})$, which allows the description of the problem 
in terms of continuous
frequencies. Because the model we study is 
isotropic, the damping matrix is diagonal $\gamma^{\mu\nu}(t)=\gamma(t) 
\delta^{\mu\nu}$,  and $\gamma (t)$ is obtained directly from the first 
derivative $\dot{\epsilon}(t)$.   
After introducing the substitution 
$\xi= (\omega+\omega')/2,\zeta=\omega-\omega'$, 
and considering only  processes for which $\zeta<<1$, we find that
$\gamma(t)=\bar{\gamma}(T)\delta(t)$ with 
 \begin{eqnarray}
\label{sgama}
\bar{\gamma}(T)&=&-\frac{2\pi\hbar}{M}\int_{0}^{\infty} d\xi 
\varphi(\xi)\frac{\partial N(\xi)}{\partial\xi}, \\ \nonumber
\varphi(\xi) &=& \frac{\xi^2}{2c^2} \sum_{n = 1}^{\infty} \sin^2 
(\delta_{n + 1} - \delta_n).
\end{eqnarray}

The phase shifts $\delta_{n}$ of the eigenfunctions $\eta_{nm}$ can be
calculated using the Fredholm method \cite{amir,vlad}. Their evaluation
requires the computation of the matrix elements $\mathcal{A}
_n=\langle n|V|n\rangle$, because 
$\delta_n=\arctan[\pi\mathcal{A}_n/(1+\mathcal{A}_n)]$.
 Here $|n \rangle$ are the eigenstates of the unperturbed
operator in Eq.\ (\ref{etaeq}).  In
 the limit of a small vortex-antivortex separation, $d \ll 
r$,
 the potential reads $V(r)=d^2/[4(r^2+d^2/4)^2]$. One then obtains
 $$ \mathcal{A}_n(\xi)=-\pi\xi \partial_\xi \left[I_n(\xi
 d/2c)K_n(\xi d/2c)\right], $$
where $I_n, K_n$ are the modified Bessel
 functions of the first and the second kinds, respectively.
Using the Drude model and assuming that the charge carriers
are attached
to the defects, we can relate $\bar{\gamma}$ to the inverse of the mobility, 
$\mu^{-1} = M \bar{\gamma} / e $. 
We then find 
\begin{eqnarray}
\mu^{-1}&=&\frac{\pi\hbar\beta}{e c^2}\sum_{n=1}^{\infty}\int_0^\infty d\xi \frac{\xi^2 e^{\beta\xi}}{(e^{\beta\xi}-1)^2}\frac{\mathcal{B}_n^2(\xi)}{[1+\mathcal{B}_n^2(\xi)]},\\ \nonumber 
\mathcal{B}_n &=& \frac{\pi(\mathcal{A}_{n+1}-\mathcal{A}_n)}{
1+\mathcal{A}_n+\mathcal{A}_{n+1}+(1+\pi^2)\mathcal{A}_n
\mathcal{A}_{n+1}}.
\end{eqnarray}

This expression leads to a vanishing $\mu^{-1}$ as $T \to 0$ because
in this limit there are no magnons to scatter the topological defects
and the latter behave as free particles. However, this limit will never
be reached in systems like LSCO because at $T_f \sim 30$K the dipoles
freeze and the charge becomes localized, leading to an upturn of the
resistivity. At high $T$ we obtain
\begin{equation}
\label{htl}
\mu^{-1}(T)=\frac{\pi k_B T}{e c^2}\sum_n\int_{0}^{\infty}
d\xi \frac{\mathcal{B}_n^2(\xi)}{1+\mathcal{B}_n^2(\xi)},   
\end{equation} 
which indicates that the resistivity $\rho=(\mu n_h e)^{-1}$ varies linearly with $T$, as observed 
experimentally \cite{ando,takagi}. Here $n_h$ denotes the
 charge carriers concentration. This behavior is expected to
 hold for
$T < T_{KT}$, where $ T_{KT}$ is the Kosterlitz-Thouless temperature, 
at which the vortex anti-vortex pairs will eventually
 unbind. An attempt
to roughly estimate $ T_{KT}$ has provided $ T_{KT} \sim J S^2$ \cite{nils}. 
Using that $J \sim 0.1$eV and $S = 1/2$,  
we find $T_{KT} \sim 300$K. Despite of the crudeness of the estimates,
this result compares quite well with the experimental data \cite{takagi}, 
which shows deviations from linearity at $T \sim 400$K.
Given that $ T_{KT}$ represents the temperature at which few pairs start to
unbind, it is not surprising that the linear behavior can hold
up to higher T. 

The order of magnitude of the resistivity can also be promptly evaluated from our calculations. Although the exact microscopic values of the parameters $d$ and $c$
are not available, we nevertheless provide an
estimate of these. Taking $\hbar c/ak_B \sim T_{KT}$ and 
$d \sim a$, we find $\mu^{-1} \sim 0.1$Vs/cm$^2$ for $T = 200$K, in good 
agreement with Fig.\ 2 of Ref.\ \cite{ando}.

In order to provide also a comparison with the experimental data concerning
the anisotropy in dc transport, we should 
include the spin-wave anisotropy
in the spiral state. The analytical 
calculations which led to Eqs.\ (\ref{sgama}) -
(\ref{htl}) cannot be performed for an anisotropic 
magnon bath. Nevertheless, we provide a rough estimate 
of this effect by considering the 
dependence of the inverse mobility in 
Eq.\ (\ref{htl}) on the spin-wave
velocity $c$. Recalling that $\mathcal{B}_n$ is a 
function of $\xi d/2c$, we find that 
$\mu^{-1}\propto c^{-1}$. One could then expect that in 
the original anisotropic system  
$\mu^{-1}_\perp\propto c_\perp^{-1}$ and 
$\mu^{-1}_\parallel\propto c_\parallel^{-1}$. 
This result can also be understood on physical grounds: 
since dissipation is provided by the excitations of the magnon bath,
spin-waves with higher velocity are less effective in scattering
the defects. As a consequence, the resistivity in the spiral direction 
is larger than that in the direction perpendicular to it,  
$\gamma_{\perp}/\gamma_{\parallel}= \rho_{\perp}/{\rho}_{\parallel} = 
c_\parallel/c_\perp=\sqrt{\cos{(Qa)}}$. For $x=0.04$, we find
$\sqrt{\cos{(Qa)}}=0.98$, in agreement with
transport measurements by Ando {\it et al.} \cite{ando}, which found 
that for $100$K$<T<200$K the resistivity along the $a$-axis in LSCO is 
slightly smaller than
along the $b$-axis (in orthorhombic coordinates). We note that the IC peaks
observed in neutron scattering correspond to the $b$-direction, which coincides
with the spiral axis because the breaking of translational symmetry within the
spiral picture has its origin in the spiral chirality (which can rotate 
clock- or counter-clockwise). Thus, $\rho_\perp \equiv \rho_a, \rho_\parallel 
\equiv \rho_b$
and $\rho_a < \rho_b$, as experimentally observed. This result indicates
that the anisotropy measured in the SG phase does
{\it not} provide evidence for the existence of diagonal stripes, but instead,
is also the result expected within a more realistic, albeit more complex,
model which does not need to  appeal to charge order at such small doping
concentrations. 

In conclusion, we propose a description of the transport properties in the SG
phase of cuprates based on the dissipative dynamics of topological defects in 
a spiral state. Using the collective-coordinate method, we derive the
effective action of the topological defects, which are coupled to a bath of 
magnons.
The scattering of magnons by the potential provided by the topological
defects leads to a dissipative motion for these defects. Under the assumption 
that the holes are attached to the defects, the corresponding
damping matrix is calculated, and is related to the in-plane resistivity. Its
$T$ dependence and anisotropic behavior are in agreement with the
available experimental data, indicating that further investigations are 
required to distinguish between spiral spin states and diagonal stripes 
in the SG phase of cuprates. 

We acknowledge fruitful discussions with N.\ Hasselmann, A.\ H.\
Castro Neto, V.\ Gritsev, and A.\ Villares Ferrer. One of us (AOC) wishes to 
thank Funda\c{c}\~{a}o de Amparo \`{a} Pesquisa no Estado de S\~{a}o Paulo 
(FAPESP) and Conselho Nacional de Desenvolvimento Cient\'{\i}fico e 
Tecnol\'{o}gico (CNPq) for their financial aid.
This work was mainly supported by the Swiss
National Foundation for Scientific Research under grant No. 620-62868.00. 


\end{document}